\documentclass[11pt]{article}
\usepackage{amsmath,amssymb,color,graphics,epsfig,cite}
\usepackage{graphicx,subfigure}
\usepackage{xcolor}
\usepackage[normalem]{ulem}

\usepackage{amsfonts}

\newcommand{\hoch}[1]{$\, ^{#1}$}

\newcommand{\be}{\begin{equation}}
	\newcommand{\ee}{\end{equation}}
\newcommand{\bea}{\setlength\arraycolsep{2pt} \begin{eqnarray}}
	\newcommand{\eea}{\end{eqnarray}}
\newcommand{\nn}{\nonumber}

\def\fft#1#2{{\frac{#1}{#2}}}

\def\0{{\sst{(0)}}}
\def\1{{\sst{(1)}}}
\def\2{{\sst{(2)}}}
\def\3{{\sst{(3)}}}
\def\4{{\sst{(4)}}}
\def\5{{\sst{(5)}}}
\def\6{{\sst{(6)}}}
\def\7{{\sst{(7)}}}
\def\8{{\sst{(8)}}}
\def\sst#1{{\scriptscriptstyle #1}}

\begin{document}
	
\begin{center}

{\Large {\bf Lorentz-Violating (Regular) Black Holes in Einstein Gravity}}
		
\vspace{20pt}

Zhi-Chao Li\hoch{1} and H. L\"u\hoch{1,2}

\vspace{10pt}

{\it \hoch{1}Center for Joint Quantum Studies, Department of Physics,\\
			School of Science, Tianjin University, Tianjin 300350, China }

\medskip

{\it \hoch{2}The International Joint Institute of Tianjin University, Fuzhou,\\ Tianjin University, Tianjin 300072, China}

\vspace{40pt}
		
\underline{ABSTRACT}
\end{center}

We introduce a minimally coupled dark sector based on a nonlinear electrodynamics
to generate regular spacetime textures that interpolate from a regular core to an asymptotic Lorentz-violating conical geometry. This construction provides a simple mechanism, in Einstein gravity with minimally coupled matter, for obtaining Lorentz-violating Schwarzschild and Reissner-Nordstr\"om black holes. The framework further allows us to construct Lorentz-violating regular black holes, including the Bardeen and Hayward black holes, as well as a class of electrically-charged regular black holes.

\vfill {\footnotesize lizc@tju.edu.cn\ \ \ mrhonglu@gmail.com}
	
\thispagestyle{empty}
\pagebreak
\newpage

\section{Introduction}
\label{sec:intro}

Lorentz symmetry is a cornerstone of both general relativity and the Standard Model of particle physics. Possible violations of local Lorentz invariance can be systematically described in the gravitational sector of the Standard-Model Extension \cite{Kostelecky:2003fs}. Bumblebee gravity provides a simple dynamical realization, in which a vector field acquires a nonzero vacuum expectation value and spontaneously breaks local Lorentz symmetry \cite{Bluhm:2004ep}. Static black-hole solutions in this theory are known to exhibit Lorentz-violating conical asymptotics \cite{Casana:2017jkc,Maluf:2020kgf}. However, this apparent simplicity comes with a strong rigidity. Once the bumblebee vector has only the radial component and is constrained to have a fixed norm, its field equation imposes nontrivial restrictions on the geometry and on the matter sector. As a result, the coupling between bumblebee gravity with matter theories, such as Maxwell theory, is technically cumbersome and has to be constructed in a case-by-case basis \cite{Liu:2024dei,Chen:2025ypx,Li:2025tcd,Ovcharenko:2026rvj}. The difficulty is exacerbated for the more complicated case of nonlinear electrodynamics (NLED) \cite{Li:2026tae}.

In this paper we take a very different approach. Instead of deriving the
Lorentz-violating cone (LVC) from the complicated non-minimally coupled bumblebee gravity, we realize it directly in Einstein gravity through a minimally coupled dark sector, which we assume decouples or weakly couples with ordinary matter. The dark sector, which can be described by a suitable NLED, supports a regular Lorentz-violating spacetime texture (LVT) that interpolates from a regular core to an asymptotic LVC. The salient feature of this construction is its modularity. The dark sector supports the LVT which is massless, while ordinary black hole matter can be added independently: any static spherically-symmetric spacetime supported by matter sector satisfying \(\rho^{\rm mat}+p_r^{\rm mat}=0\) can be embedded into the texture. This includes the celebrated Schwarzschild and Reissner-Nordstr\"om (RN) black holes, and a variety of black holes supported by general classes of NLED, which makes the construction of Lorentz-violating regular black holes a simple task, while the similar attempt in bumblebee gravity meets only limited success \cite{Li:2026tae}.

The paper is organized as follows. In Sec.~2 we analyse the LVC and construct its regularized LVT in Einstein gravity with minimally coupled dark sector. We propose the suitable NLED for the corresponding dark sector. In Sec.~3 we embed standard and regular static spherically-symmetric black holes into the LVC or LVT, and show that the black hole thermodynamics follow straightforwardly. In Sec.~4 we include a cosmological constant and discuss the resulting asymptotic structure. We conclude the paper in Sec.~5.

\section{Regular Lorentz-violating spacetimes}

The class of Lorentz-violating backgrounds we consider is static and spherically symmetric, with metric
\be
ds^2= -dt^2 + (1 + \ell) dr^2 + r^2 d\Omega_2^2\,.
\label{LVcone}
\ee
The spacetime is simply a direct product of time and a three-dimensional Euclidean cone; we therefore refer it as LVC. LVC arises naturally in bumblebee gravity \cite{Casana:2017jkc}, which makes the theory an important framework to study Lorentz violation in gravity. However, its deficiency discussed in the Introduction prompts us to ask whether such a solution can exist in Einstein gravity with minimally coupled matter. To this end, it is necessary first to check the energy condition of the supporting matter.
For $T^\mu{}_\nu = {\rm diag}\{-\rho,p_r,p_\perp,p_\perp\}$, we have
\be
\{\rho,p_r,p_\perp\} = \fft{\ell}{(1+\ell)r^2} \{1,-1,0\}\,.
\ee
Thus, for $\ell>0$, the matter source satisfies {\it all} the major energy conditions, implying the LVC of \eqref{LVcone} is perfectly consistent within Einstein gravity. However, this cone suffers from a curvature power-law singularity at the center $r=0$, with the Ricci scalar $R=2\ell/((1+\ell)r^2)$.

In order to resolve the conical singularity, we construct gravitational textures, which are regular spacetimes devoid of mass or angular momentum. A simple such LVT example is
\be
ds^2 = -(1+\ell) f(r) dt^2 + \fft{dr^2}{f(r)} +
r^2 d\Omega_2^2\,,
\qquad
f=f_{\rm T}\equiv
\fft{1}{1+\ell}
\left(
1+\fft{a^2\ell}{r^2+a^2}
\right).
\label{texture1}
\ee
The metric is asymptotic to LVC,  with a regular de Sitter (dS) core at the center.
Specifically, we have
\be
f_{\rm T}=
1-\fft{\ell}{1+\ell}\fft{r^2}{a^2}
+{\cal O}(r^4)\,,
\qquad\hbox{and}\qquad
f_{\rm T}
=
\fft{1}{1+\ell}
+
\fft{a^2\ell}{(1+\ell) r^2}
+
{\cal O}(r^{-4})\,.
\ee
The metric is massless, since the leading falloff is $1/r^2$. The parameter $a$ characterizes the size of the texture. The texture becomes the singular LVC when $a=0$, and it reduces to the standard Minkowski spacetime up to a constant time rescaling in the limit $a\rightarrow \infty$. The matter energy-momentum tensor supporting this spacetime texture is given by
\be
\rho
=
\fft{\ell(r^2+3a^2)}
{(1+\ell)(r^2+a^2)^2}
=
-p_{r}\,,
\qquad
p_{\perp}
=
\fft{\ell a^2(r^2-3a^2)}
{(1+\ell)(r^2+a^2)^3}\,.
\ee
It can be checked that the matter source satisfies the dominant-energy condition (DEC), while the strong-energy condition (SEC) is violated near the core, i.e.~$r^2<3a^2$.

We have thus shown that regular and horizonless LVTs can be constructed in Einstein gravity with suitable energy conditions. We now consider matter candidates in Lagrangian mechanics, so that the total Lagrangian density is
\be
{\cal L}_{\rm tot}
=
\sqrt{-g} \big(R + L_{\rm DS}\big)\,.
\ee
We shall assume that $L_{\rm DS}$ does not interact directly (or only weakly interacts) with any observable matter, and hence we refer to it as the dark sector. To support the LVC \eqref{LVcone}, a simple candidate involves a dark Maxwell field with a non-dynamical scalar, with the Lagrangian
\be
L_{\rm DS} =
-2\phi^{-1}\Big({\cal F}+\fft{1}{4\alpha}\Big)\,,
\qquad
{\cal F}=\frac{1}{4}F^{\mu\nu} F_{\mu\nu}\,.
\label{dselec1}
\ee
Here, the coupling constant $\alpha$ has dimension of length squared, so that $\alpha {\cal F}$ is dimensionless. The scalar equation of motion implies a constant constraint on ${\cal F}$,
namely
\be
{\cal F}=-\fft{1}{4\alpha}\,,
\ee
which is the source for Lorentz violation. For the metric \eqref{LVcone}, the scalar and Maxwell field are
\be
A= \sqrt{\fft{1+\ell}{2\alpha}}\, r\, dt\,,
\qquad
\phi = \fft{1+\ell}{2\alpha \ell} r^2\,.
\ee
In other words, the LVC is supported by a constant radial component of the dark electric field. Note that the magnetic dual of \eqref{dselec1} can be equivalently expressed as a specific NLED, namely
\be
L_{\rm DS} = -2\Big(\phi\, {\cal F} + \fft{1}{4\alpha \phi}
\Big)\,,
\qquad
\rightarrow
\qquad
L_{\rm DS} = - \fft{2}{\alpha}\sqrt{\alpha {\cal F}}\,.
\label{dsmag1}
\ee
This provides the flexibility to construct a Lagrangian in the dark sector for the more general class of spacetime textures. For example, the NLED that supports the texture geometry \eqref{texture1} is given by
\be
L_{\rm DS}
=
-\fft{2}{\alpha}\sqrt{\alpha {\cal F}} \,\fft{
(1 + 3\beta\sqrt{\alpha {\cal F}})}
{(1 +\beta \sqrt{\alpha {\cal F}})^2}\,,
\ee
where $\beta$ is a dimensionless parameter. The Lagrangian reduces to \eqref{dsmag1} when $\beta=0$ and it vanishes in the $\beta\rightarrow \infty$ limit. This NLED gives rise to the texture geometry \eqref{texture1} provided that the dark Maxwell field is magnetic, namely $F_{\theta\phi}=p\sin\theta$, with ${\cal F}=p^2/(2r^4)$ and the parameters $(a,p)$ determined to be
\be
a^2=\fft{\alpha \beta \ell}{1+\ell}\,,\qquad p=\fft{\sqrt{2\alpha}\,\ell}{1+\ell}\,.
\label{apcons}
\ee
It is worth emphasizing that this construction is not merely a matching of the $tt$ component of the Einstein equations. With our convention, a purely magnetic NLED sector described by $L({\cal F})$ gives
\be
\rho=-\frac{1}{2}L\,,
\qquad
p_r=\frac{1}{2}L\,,
\qquad
p_\perp=\frac{1}{2}L-{\cal F} L_{\cal F}\,.
\ee
The conservation of the energy-momentum tensor, i.e.~$p_\perp=-\rho-\fft{r}{2}\rho'$, agrees with the angular component of the Einstein equations for the texture geometry.

To show that the above texture is not an isolated example, we record two more examples. It is instructive to write the metric profile as
\be
f_{\rm T}(r) =\fft{1}{1+\ell} +  \frac{\ell}{1+\ell} U(r)\,,
\label{general_texture_profile_main}
\ee
the two examples are
\bea
U_1 &=& \left(
1+\frac{r^2}{a^2}\right)^{-n}, \qquad n\ge 1\,,\qquad U_2 = \exp\left(-\frac{r^2}{a^2}\right);\nn\\
L_1 &=&
-\frac{2}{\alpha}\sqrt{\alpha{\cal F}}
\left[1-\left(1+\frac{1}{\beta\sqrt{\alpha{\cal F}}}\right)^{-n}
+\frac{2n}{\beta\sqrt{\alpha{\cal F}}}\left(1+\frac{1}{\beta\sqrt{\alpha{\cal F}}}
\right)^{-n-1}\right],\nn\\
L_2 &=&  -\frac{2}{\alpha}\sqrt{\alpha{\cal F}}\left[1-\exp\left(
-\frac{1}{\beta\sqrt{\alpha{\cal F}}}\right)+
\frac{2}{\beta\sqrt{\alpha{\cal F}}}
\exp\left(-\frac{1}{\beta\sqrt{\alpha{\cal F}}}\right)\right],
\eea
with the parameter constraint \eqref{apcons}. In both examples, we have $U(0)=1$, $U'(0)=0$ and $\lim_{r\rightarrow\infty}rU(r)=0$, and hence they describe massless regular textures that approach the LVC at large $r$. It can be checked that both examples satisfy DEC, but not SEC.

We now follow the analogous argument of \cite{Li:2023yyw} to show that the LVTs necessarily violate the SEC, which requires
\be
\rho + p_\perp = -{\cal F} L_{\cal F}\ge 0\,,\qquad p_\perp \ge 0\,.
\ee
For the regular LVT, we have a finite $\rho(0)$ and vanishing $\rho(\infty)$. The quantity \({\cal F}=p^2/(2r^4)\) runs from 0 to infinity as $r$ runs from infinity to zero. Thus we must have $L(0)=0$ and $L(\infty)$ is finite so that ${\cal F} L_{\cal F}$ vanishes at the core. On the other hand, the above NEC implies that $L_{\cal F}$ must be negative, indicating that $L(\infty)$ at the core must be negative. Thus, the SEC must be violated at the core.

\section{(Regular) black holes}

Having constructed the LVT supported by an NLED in the dark sector, we now study how mass and standard matter affect the spacetime. First, the texture allows a straightforward Schwarzschild-type mass deformation, namely
\be
f\rightarrow f= f_{\rm T} - \fft{2m}{r}\,.
\ee
Here $f_{\rm T}$ is the metric profile of the texture discussed in the previous section. Using the texture \eqref{texture1} as a concrete example, the spacetime describes a Lorentz-violating black hole provided that the metric function has a positive root. The horizon $r_+$ is determined by $f(r_+)=0$. The mass, temperature and entropy can be calculated in the standard manner, given by
\be
M=\sqrt{1+\ell}\, m\,,
\qquad
T=
\fft{\sqrt{1+\ell}}{4\pi}f'(r_+)\,,
\qquad
S=\pi r_+^2\,.\label{MTS}
\ee
Treating the texture parameters $(\ell,a)$ as thermodynamic constants,
we obtain the first law $dM=TdS$ for the Schwarzschild-type deformation. Since $m=\frac{1}{2} r_+ f_{\rm T}(r_+)$, and $f_{\rm T}(0)$ is finite, there is no mass gap for the Schwarzschild-type black holes. The simplicity should be contrasted to bumblebee gravity, where the Wald entropy formula is not compatible with the first law and has to be modified \cite{Chen:2025ypx, Li:2025tcd}.

For the usual Maxwell and Born--Infeld matters \cite{Born:1934gh,Li:2016nll},
the metric profiles of the dyonic black holes become simply,
\bea
f_{\rm Max} &=& f_{\rm T} - \fft{2m}{r} + \fft{4(Q^2+P^2)}{r^2}\,,\nn\\
f_{\rm BI} &=& f_{\rm T} - \fft{2m}{r} + \fft{r^2}{6 \alpha_{\rm BI} ^2}
\Big[
1- {}_2F_1\Big(-\fft34,-\fft12;\fft14;-\fft{16 \alpha_{\rm BI} ^2(Q^2+P^2)}{r^4}\Big)
\Big]\,.
\eea
These are Lorentz-violating generalizations of the RN and Born-Infeld
black holes. Owing to the Lorentz-violating parameter $\ell$, the mass, temperature and entropy are given by \eqref{MTS}. The electric and magnetic charges remain $(Q,P)$, with the corresponding thermodynamic potentials scaled by $\sqrt{1+\ell}$. The first law of black hole thermodynamics can be easily verified.

Generally speaking, if we have a matter system satisfying $\rho^{\rm mat}+p_r^{\rm mat}=0$, then it preserves the texture's geometric structure
of $g_{tt}g_{rr}=$constant. The metric function is simply modified to $f=f_{\rm T}+f^{\rm mat}$ where $f^{\rm mat}$ satisfies a simple first-order differential equation
\be
-r \fft{d}{dr} f^{\rm mat} - f^{\rm mat} = r^2\rho^{\rm mat}\,.
\label{fmat}
\ee
This follows directly from the $tt$ component of the Einstein equations for the metric ansatz above.
Note that this equation is independent of the detailed form of the texture geometry, and applies equally to asymptotically flat, conical, and (A)dS-type backgrounds.
Our mechanism thus allows all the static black holes, satisfying $\rho^{\rm mat}+p_r^{\rm mat}=0$, to be embedded into the LVT.

One application is to construct regular black holes asymptotic to LVC.
For examples, the matter energy-momentum tensors of the Bardeen and Hayward black holes
\cite{Bardeen:1968,Hayward:2005gi} are respectively given by
\bea
{\rm Bardeen:}\qquad \rho^{\rm mat} &=& - p_r^{\rm mat}
= \fft{6g^2 m_0}{(g^2 + r^2)^{\fft52}}\,,
\qquad
p_\perp^{\rm mat}
= \fft{3g^2 m_0 (3r^2 - 2g^2)}{(g^2 + r^2)^{\fft72}}\,;
\nn\\
{\rm Hayward:}\qquad \rho^{\rm mat} &=& - p_r^{\rm mat}
= \fft{6g^3 m_0}{(g^3 + r^3)^2}\,,
\qquad
p_\perp^{\rm mat}
= \fft{6g^3 m_0 (2r^3-g^3)}{(g^3 + r^3)^3}\,.
\eea
For these exact same energy-momentum tensors, the most general solutions to \eqref{fmat} are respectively given by
\be
f_{\rm B}
=
f_{\rm T}
- \fft{2(m-m_0)}{r}
-\fft{2m_0\, r^2}{(g^2 + r^2)^{\fft32}}\,,
\qquad
f_{\rm H}
=
f_{\rm T}
- \fft{2(m-m_0)}{r}
-\fft{2m_0\, r^2}{g^3 + r^3}\,.
\ee
Here $m$ is the free Schwarzschild-type mass parameter. When $m=m_0$, the Schwarzschild-type pole is removed. In this case, both solutions have a regular small-$r$ expansion,
\be
f_{\rm B,H}(r)
=
1-
\left( \fft{\ell}{(1+\ell)a^2}+\fft{2m_0}{g^3}
\right)r^2
+{\cal O}(r^4)\,.
\ee
Thus, the center is regular for $m=m_0$. For $m\neq m_0$, the $1/r$ term remains and the geometry is singular at the origin. When the texture is turned off, \(f_{\rm T}=1\), and $m=m_0$, the metric reduces to the standard Bardeen or Hayward solutions, respectively.
Thus, we have constructed the Bardeen and Hayward regular black holes embedded in the LVT. Note that the regular texture geometry is essential in the construction of Lorentz-violating regular black holes, as we need not only remove the mass pole singularity but also the conical singularity of the LVC.

There are currently two main approaches to construct Lagrangian mechanics that leads to the effective energy-momentum tensors of regular black holes. One is based on pure gravity using an infinite number of quasi-topological curvature terms \cite{Bueno:2024dgm}. Such an approach typically works most naturally in higher dimensions, but it is certainly compatible with our approach in higher dimensions, as they both satisfy $g_{tt} g_{rr}=$ constant. The other is to use a suitable NLED in the form of \(L({\cal F})\) as the matter sector. It was shown in Refs.~\cite{AyonBeato:2000zs,Fan:2016hvf} that the \(L({\cal F})\) functions associated with standard regular black-hole
geometries, such as the Bardeen and Hayward metrics, are typically nonanalytic
in ${\cal F}$, and that the corresponding regular solutions are
naturally realized in the magnetic branch.
In fact, a no-go theorem has been established for electrically charged regular
black holes in single-valued \(L({\cal F})\) theories \cite{Bronnikov:2000vy}.

Recently, several extensions of the NLED approach have been proposed to overcome these limitations.
Analytical $L({\cal F})$ theories with a weak-field expansion of the form
\be
L({\cal F})
=
- {\cal F}+\alpha_1 {\cal F}^2+\alpha_2 {\cal F}^3+\cdots
\ee
can be constructed to give rise to regular black holes satisfying DEC \cite{Li:2023yyw}. This analyticity condition rules out both Bardeen and Hayward black holes. (We do not impose the analyticity condition on the dark sector, as it is hidden from our direct observation.) Electrically charged regular black holes can also be constructed using an auxiliary Maxwell-scalar formulation of NLED with a non-dynamical scalar \cite{Li:2024vua}. Specifically, we consider the matter Lagrangian of the form
\be
L_{\rm MS} = - \phi^{-1}{\cal F} - V(\phi)\,,
\ee
for a suitable scalar potential of the auxiliary scalar. This formalism is fully consistent with our LVT. As a simple explicit example, we take the scalar potential to be \cite{Li:2024vua}
\be
V(\phi) = \fft{1}{\alpha_{\rm MS}} \left(1-\sqrt{\phi}\right)^2\,.
\ee
It is then straightforward to construct the electrically charged black hole embedded in the LVT, given by
\bea
f(r)
&=&
f_{\rm T}(r)
-\fft{2m}{r}
+\fft{q^2}{4r^2}\,
{}_2F_1\left(
\fft14,1;\fft54;
-\fft{\alpha_{\rm MS} q^2}{2r^4}
\right),
\nn\\
F_{0r}
&=&
\fft{\sqrt{1+\ell}\,q\,\phi(r)}{r^2}\,,
\qquad
\phi(r)
=
\left(1+\fft{\alpha_{\rm MS}q^2}{2r^4}\right)^{-2}.
\eea
For sufficiently large $m$, the solution describes
a black hole of mass $M= m \sqrt{1+\ell}$ and electric charge $Q_e=q/4$. The first law can be easily checked. As in the previous simpler Bardeen and Hayward examples, the central Schwarzschild-type pole can be removed by imposing the mass/charge relation
\be
m =\frac{\pi |q|^{3/2}} {16\,2^{1/4}\alpha_{\rm MS}^{1/4}}\,.
\ee
This leads to a Lorentz-violating regular black hole carrying electric charges. When the texture profile is $f_{\rm T}=1$, the solution reduces to the regular black hole in the asymptotic Minkowski spacetime. Several such electrically charged regular black holes were constructed in \cite{Li:2024vua}, and they can all be embedded in the LVT, without sacrificing the black hole regularity.

\section{Including the cosmological constant}

The energy-momentum tensor associated with the cosmological constant $\Lambda$
also has the property $\rho + p_r=0$, and hence it is fully consistent with our texture geometry. The effect of including the cosmological constant simply adds an extra term to $f$, i.e.
\be
f = -\fft13\Lambda\, r^2 + f_{\rm T}+f^{\rm mat}\,.
\ee
The cosmological constant alters the asymptotic structure, and the leading large-\(r\) behavior is governed by the cosmological term rather than by the LVC. It is now more natural to adopt the canonical (A)dS time normalization and set $g_{tt} g_{rr}=-1$, so that the asymptotic structure becomes
\be
-g_{tt}=f(r)= -\fft{\Lambda}{3}r^2 + \fft{1}{1+\ell} - \fft{2m}{r} +\cdots\,.
\ee
Thus, the dark-sector texture leaves a conical imprint through the constant term \(1/(1+\ell)\), while the spacetime is asymptotically locally (A)dS rather than asymptotic to (A)dS or the LVC. It was observed that a negative cosmological constant can be viewed as the pressure \(P=-\Lambda/(8\pi)\) in extended black-hole thermodynamics \cite{Kastor:2009wy}. For the $g_{tt}g_{rr}=-1$ convention adopted here, the corresponding thermodynamic volume is simply $V_{\rm th}=4\pi r_+^3/3$, independent of the texture parameter $\ell$. If instead one had kept the previous conical time normalization, \(g_{tt}=-(1+\ell)f\), the thermodynamic volume would be rescaled to \(V_{\rm th}=4\pi\sqrt{1+\ell}\,r_+^3/3\).

\section{Conclusion}

In this paper, we showed that the LVC and its regularized texture are perfectly consistent with Einstein gravity, with the former satisfying all major energy conditions, whereas the latter satisfies all but the SEC. We then introduce a minimally coupled dark sector in the form of NLED that gives explicit realization of these Lorentz-violating spacetimes in Lagrangian mechanics.

This formalism allows us to embed a variety of previously known black holes in these Lorentz-violating spacetimes, including the Schwarzschild and RN black holes. Furthermore, our regular textures make it straightforward to embed regular black holes such as the Bardeen, Hayward and the previously constructed electrically-charged regular black holes in our Lorentz-violating backgrounds without sacrificing the spacetime regularity.

The present construction gives a simple realization of Lorentz-violating (regular) black holes with conical asymptotics in Einstein gravity. Its compatibility with (minimally) coupled matter is far superior to bumblebee gravity, which requires convoluted and case-by-case matter couplings to achieve the same spacetime geometry. The new black hole solutions can be used as controlled analytical backgrounds for isolating the effects of Lorentz-violating asymptotics on thermodynamics, geodesic motion, shadows, lensing and quasi-normal modes, without the additional complications introduced by convoluted non-minimal gravitational couplings. The regular solutions further provide a controlled setting for studying how a minimally coupled dark sector can resolve central singularities and how such a resolution affects strong-field physics.

\section*{Acknowledgements}

This work is supported in part by the National Natural Science Foundation of China (NSFC) grants No.~12375052 and No.~11935009, and also by the Tianjin University Self-Innovation Fund Extreme Basic Research Project Grant No.~2025XJ21-0007.


\begin{thebibliography}{99}

\bibitem{Kostelecky:2003fs}
	V.A.~Kosteleck\'y,
	``Gravity, Lorentz violation, and the standard model,''
	Phys. Rev. D {\bf 69}, 105009 (2004),
	arXiv:hep-th/0312310.
	
\bibitem{Bluhm:2004ep}
	R.~Bluhm and V.A.~Kosteleck\'y,
	``Spontaneous Lorentz violation, Nambu-Goldstone modes, and gravity,''
	Phys. Rev. D {\bf 71}, 065008 (2005),
	arXiv:hep-th/0412320.
	
\bibitem{Casana:2017jkc}
	R.~Casana, A.~Cavalcante, F.P.~Poulis and E.B.~Santos,
	``Exact Schwarzschild-like solution in a bumblebee gravity model,''
	Phys. Rev. D {\bf 97}, 104001 (2018),
	arXiv:1711.02273 [gr-qc].
	
\bibitem{Maluf:2020kgf}
	R.V.~Maluf and J.C.S.~Neves,
	``Black holes with a cosmological constant in bumblebee gravity,''
	Phys. Rev. D {\bf 103}, 044002 (2021),
	arXiv:2011.12841 [gr-qc].
	
\bibitem{Liu:2024dei}
J.Z.~Liu, W.D.~Guo, S.W.~Wei and Y.X.~Liu,
``Charged spherically symmetric and slowly rotating charged black hole solutions in bumblebee gravity,''
	Eur. Phys. J. C {\bf 85}, no.~2, 145 (2025),
	arXiv:2407.08396 [gr-qc].

\bibitem{Chen:2025ypx}
Y.Q.~Chen and H.S.~Liu,
``Taub-NUT-like black holes in Einstein-bumblebee gravity,''
Phys. Rev. D \textbf{112}, no.8, 084040 (2025)
doi:10.1103/wmhj-83x3
[arXiv:2505.23104 [gr-qc]].

\bibitem{Li:2025tcd}
S.~Li, L.~Liang and L.~Ma,
``Dyonic RN-like and Taub-NUT-like black holes in Einstein-bumblebee gravity,''
JCAP \textbf{03}, 005 (2026)
doi:10.1088/1475-7516/2026/03/005
[arXiv:2510.04405 [gr-qc]].

\bibitem{Ovcharenko:2026rvj}
H.~Ovcharenko,
``Exact Kerr-Newman-(A)dS and other spacetimes in bumblebee gravity: employing a simple generating technique,''
[arXiv:2601.16037 [gr-qc]].

	\bibitem{Li:2026tae}
	Z.C.~Li and H.~L\"u,
	``When Bumblebee Meets NLED: Lorentz-Violating Black Holes and Regular Spacetimes,''
	arXiv:2605.13963 [gr-qc].

	\bibitem{Li:2023yyw}
	Z.C.~Li and H.~L\"u,
	``Regular black holes from analytic \(f(F^2)\),''
	Eur. Phys. J. C {\bf 83}, 755 (2023),
	arXiv:2303.16924 [gr-qc].
	
	\bibitem{Born:1934gh}
	M.~Born and L.~Infeld,
	``Foundations of the new field theory,''
	Proc. Roy. Soc. Lond. A {\bf 144}, 425 (1934).
	
	\bibitem{Li:2016nll}
	S.~Li, H.~L\"u and H.~Wei,
	``Dyonic (A)dS black holes in Einstein-Born-Infeld theory in diverse dimensions,''
	JHEP {\bf 07}, 004 (2016),
	arXiv:1606.02733 [hep-th].
	
	\bibitem{Bardeen:1968}
	J.M.~Bardeen,
	``Non-singular general-relativistic gravitational collapse,''
	in {\it Proceedings of the International Conference GR5},
	Tbilisi, U.S.S.R. (1968).
	
	\bibitem{Hayward:2005gi}
	S.A.~Hayward,
	``Formation and evaporation of nonsingular black holes,''
	Phys. Rev. Lett. {\bf 96}, 031103 (2006),
	arXiv:gr-qc/0506126.
	
	\bibitem{Bueno:2024dgm}
	P.~Bueno, P.A.~Cano and R.A.~Hennigar,
	``Regular black holes from pure gravity,''
	Phys. Lett. B {\bf 861}, 139260 (2025),
	arXiv:2403.04827 [gr-qc].
	
	\bibitem{AyonBeato:2000zs}
	E.~Ay\'on-Beato and A.~Garc\'{\i}a,
	``The Bardeen model as a nonlinear magnetic monopole,''
	Phys. Lett. B {\bf 493}, 149 (2000),
	arXiv:gr-qc/0009077.
	
	\bibitem{Fan:2016hvf}
	Z.Y.~Fan and X.~Wang,
	``Construction of regular black holes in general relativity,''
	Phys. Rev. D {\bf 94}, 124027 (2016),
	arXiv:1610.02636 [gr-qc].
	
	\bibitem{Bronnikov:2000vy}
	K.A.~Bronnikov,
	``Regular magnetic black holes and monopoles from nonlinear electrodynamics,''
	Phys. Rev. D {\bf 63}, 044005 (2001),
	arXiv:gr-qc/0006014.
	
	\bibitem{Li:2024vua}
	Z.C.~Li and H.~L\"u,
	``Regular electric black holes from Einstein-Maxwell-scalar gravity,''
	Phys. Rev. D {\bf 110}, 104046 (2024),
	arXiv:2407.07952 [gr-qc].
	
	\bibitem{Kastor:2009wy}
	D.~Kastor, S.~Ray and J.~Traschen,
	``Enthalpy and the mechanics of AdS black holes,''
	Class. Quant. Grav. {\bf 26}, 195011 (2009),
	arXiv:0904.2765 [hep-th].

\end{thebibliography}
\end{document}